# Serial Symmetrical Relocation Algorithm for the Equal Sphere Packing Problem†

## WenQi Huang, Liang Yu[*]

College of Computer Science, HuaZhong univ. of Science and Technology, Wuhan 430074, China.

**Abstract**

For dealing with the equal sphere packing problem, we propose a serial symmetrical relocation algorithm, which is effective in terms of the quality of the numerical results. We have densely packed up to 200 equal spheres in spherical container and up to 150 equal spheres in cube container. All results are rigorous because of a fake sphere trick. It was conjectured impossible to pack 68 equal spheres of radius 1 into a sphere of radius 5. The serial symmetrical relocation algorithm has proven wrong this conjecture by finding one such packing.

## Introduction

How to densely pack equal spheres inside a bounded container is a problem of both theoretical and practical value. Although the problem of packing equal spheres in infinite space has been solved by (Hales, 2000), the problem of packing equal spheres inside a bounded container remains unsolved. Besides its theoretical value, it is also of practical value, because in practice all containers are bounded.

The problem of packing equal spheres in spherical containers can be used to model the treat plan of the Gamma Knife (Wang, 1999). The Gamma Knife consists of 201 cobalt-60 radioactive sources and provides an advanced approach to the treatment of tumor and vascular malformations within the head. Each radioactive source emits a gamma ray. A certain number of gamma rays of the same diameter are designedly pointed at one point in space. At this point, a spherical region (called a shot) of high radiation dose is formed. The tumor can be viewed as an approximate spherical container. All shots can be viewed as ideal equal spheres which shall be packed inside the container. No shot is allowed to extend outside the container so that no normal tissue shall be affected. No two shots should overlap with each other so that no overlapped part with too high dose shall occur in the tumor. Any good method dealing with the problem of packing equal spheres inside a spherical container is conducive to improving the treat plan of the Gamma Knife.

* Corresponding author: tel: 86-027-87543885; e-mail address: forrestyu1980@gmail.com

†This work is supported by the National Natural Science Foundation of China under Grant No.61070235; the National Natural Science Foundation of China under Grant No.61173180.

As to packing equal spheres in a cube, it has a usual application in packaging industry.

By means of a max-min optimization approach, (Maranas et al, 1995) dealt with the problem of packing $n$ equal circles in a square and obtained the quasi optimal solutions for $n \leq 30$. (Graham et al, 1998) used the billiard simulation to find out the quasi optimal packings of up to 65 equal circles in a circle. Because the equal sphere packing problem can be viewed as a 3D variant of the equal circle packing problem, we also paid due attention to the methods for packing equal circles in a bounded container.

(Gensane, 2004) has presented his best packings of up to 32 equal spheres in a cube by means of an adaptation of the billiard algorithm proposed by (Lubachevsky, 1991). (Hifi and M'Hallah, 2009) made an extensive review on the circle and sphere packing problems.

(Huang and Xu, 1999), (Huang and Kang, 2004) and (Huang et al, 2005) used the quasi physical (shortly QP) method as the local solver for packing equal or unequal circles inside a large circle. The global search strategy of (Huang and Kang, 2004) is to randomly redistribute all items when the calculation is trapped in a local optimum. The global search strategy of (Huang and Xu, 1999) is to pick the item suffering the largest total overlap and randomly relocate it when trapped.

(Huang et al, 2005) applied greedy algorithms as the global search strategy for packing unequal circles in a bounded container. Given *n* unequal circles, after having placed *i(i<n)* circles, they placed next circle by the maximum hole degree rule. The action of placing a circle of appropriate size in some certain vacant region of the container can occupy most area of this region without overlapping other circles in the container. Such placing action can produce larger hole degree than placing too large or too small circles. The greedy algorithms exploit the difference of items sizes and do not work well for equal items.

We design a serial symmetrical relocation strategy as the new global search strategy, and call our algorithm the serial symmetrical relocation algorithm.

Here we clarify two terms: configuration and packing. A configuration of *n* equal spheres is a set of locations of *n* equal spheres centers. A packing is a configuration which can meet the constraints of the equal sphere packing problem.

This paper is organized as follows. We first introduce the QP model and the QP algorithm as the local solver (denoted by A0). We propose a simple trick called the fake sphere trick in A0 to guarantee the exactness of the results. Then, we propose a serial symmetrical relocation strategy and combine it with A0 to form the serial symmetrical relocation algorithm (denoted by A1). Then, the results of up to 200 equal spheres in spherical container and up to 150 equal spheres in cube container are produced. All our results are rigorous and supported by the spheres centers locations. At last, we have made the conclusions about our work.

**The quasi physical model and method**

Let (0,0,0) be the coordinates of the container center (container radius *r0*) and *X* be the coordinates of *n* equal spheres centers (sphere radius *r*), where $X=\{X_1,...,X_n\}=\{x_1,y_1,z_1,...,x_n,y_n,z_n\}$.

We represent $\sqrt{x_i^2+y_i^2+z_i^2}$ by $|X_i|$, $\sqrt{(x_i-x_j)^2+(y_i-y_j)^2+(z_i-z_j)^2}$ by $|X_i-X_j|$. The constraints of the problem of packing *n* equal spheres in spherical container are:

$$|X_i| \leq r0-r, i=1...n \quad . \tag{1}$$

$$|X_i-X_j| \geq 2r, i,j=1..n, j\neq i \quad . \tag{2}$$

A packing is a configuration $X$ that satisfies constraints (1) and (2).

Viewing all equal spheres as light smooth elastic solids inside a rigid container, we use $d_{i0}$ to represent the $i^{th}$ sphere's deformtion caused by the container (denoted by the $0^{th}$ object). $\vec{d}_{i0}$ represents the elastic repulsion force the $i^{th}$ sphere suffers for $d_{i0}$. Its direction is from the $i^{th}$ sphere center to the container center.

$$d_{i0} = |\vec{d}_{i0}| = \begin{cases} |X_i| + r - r0, & \text{if } |X_i| + r > r0 \\ 0, & \text{else} \end{cases} \quad (3)$$

$$(i = 1,..., n)$$

We use $d_{ij}$ to represent the $i^{th}$ sphere's deformation caused by the $j^{th}$ sphere. $\vec{d}_{ij}$ represents the elastic repulsion force the $i^{th}$ sphere suffers for $d_{ij}$. Its direction is from the $j^{th}$ sphere center to the $i^{th}$ sphere center.

$$d_{ij} = |\vec{d}_{ij}| = \begin{cases} \frac{1}{2} \times (2r - |X_i - X_j|), & \text{if } |X_i - X_j| < 2r \\ 0, & \text{else} \end{cases} \quad (4)$$

$$(i, j = 1...n, j \neq i)$$

We define the potential energy of the $i^{th}$ sphere as $u_i = \sum d_{ij}^2$ $(j = 0,...n, j \neq i)$. Thus, the total potential energy of all equal spheres is defined as:

$$U(X,r,r0) = \sum_{i=1}^{n} u_i = \sum_{i=1}^{n} d_{i0}^2 + \sum_{i=1}^{n} \sum_{j=1, j \neq i}^{n} d_{ij}^2. \quad (5)$$

We call $U(X,r,r0)$ as the potential energy function of $X$. $U(X,r,r0)$ can depict the deformations of all $n$ equal spheres.

Given an initial configuration $X$, we use the quasi physical algorithm (Huang and Kang, 2004) as the local solver to locally minimize $U(X,r,r0)$. This local solver here is denoted as A0. A0 is a deterministic algorithm. Its basic idea is to simulate the elastic movement of light smooth and elastic spheres jammed in a bounded container. Such moving process is a natural locally minimizing process of the deformations of all spheres.

If $U(X,r,r0)$, which is the square sum of all spheres deformations, is reduced to zero, a packing $X$ of equal spheres of radius $r$ inside the container of radius $r0$ is obtained.

Because of the limitation of modern digital computers, we consider $U(X,r,r0)$ has approximately reached zero when it is locally minimized to less than $10^{-16}$. That means the corresponding $X$ of equal spheres of radius $r$ still has slight deformations. To get exact packings by the local solver A0, we propose a fake sphere trick.

Without loss of generality, we take standard equal sphere radius as 0.5. In our experiments, we use fake sphere of radius $0.5+10^{-8}$ instead of standard sphere.

**Theorem** $U(X,0.5+10^{-8},r0)<10^{-16} \Rightarrow U(X,0.5,r0)=0$.

**Proof:**

$$U(X,0.5+10^{-8},r0)= \sum_{i=1}^{n} d_{i0}^2 + \sum_{i=1}^{n}\sum_{j=1,j\neq i}^{n} d_{ij}^2 <10^{-16} \Rightarrow$$

$d_{i0}^2<10^{-16}$ and $d_{i0}<10^{-8}$ ($i=1,\ldots,n$), while $r=0.5+10^{-8} \Rightarrow$

$$d_{i0} = \begin{cases} \sqrt{x_i^2+y_i^2+z_i^2}+0.5+10^{-8}-r0, & \text{if } \sqrt{x_i^2+y_i^2+z_i^2}+0.5+10^{-8}>r0 \\ 0, & \text{else} \end{cases} <10^{-8} \Rightarrow$$

$(i=1,\ldots,n)$

$\sqrt{x_i^2+y_i^2+z_i^2}+0.5+10^{-8}-r0<10^{-8}$, $i=1,\ldots,n \Rightarrow$

$\sqrt{x_i^2+y_i^2+z_i^2}+0.5<r0$, $i=1,\ldots,n \Rightarrow$

$d_{i0}=0$ ($i=1,\ldots,n$), while $r=0.5$;

$$U(X,0.5+10^{-8},r0)= \sum_{i=1}^{n} d_{i0}^2 + \sum_{i=1}^{n}\sum_{j=1,j\neq i}^{n} d_{ij}^2 <10^{-16} \Rightarrow$$

$d_{ij}^2<10^{-16}$ and $d_{ij}<10^{-8}$ ($i=1,\ldots,n; j=1,\ldots i-1, i+1,\ldots,n$), while $r=0.5+10^{-8} \Rightarrow$

$$d_{ij} = \begin{cases} \frac{1}{2}\times\left(2\times(0.5+10^{-8})-\sqrt{(x_i-x_j)^2+(y_i-y_j)^2+(z_i-z_j)^2}\right), & \text{if } \sqrt{(x_i-x_j)^2+(y_i-y_j)^2+(z_i-z_j)^2}<2\times(0.5+10^{-8}) \\ 0, & \text{else} \end{cases} <10^{-8} \Rightarrow$$

$(i,j=1..n, j\neq i)$

$$\frac{1}{2}\times\left(2\times(0.5+10^{-8})-\sqrt{(x_i-x_j)^2+(y_i-y_j)^2+(z_i-z_j)^2}\right)<10^{-8} \Rightarrow$$

$\sqrt{(x_i-x_j)^2+(y_i-y_j)^2+(z_i-z_j)^2}>2\times 0.5 \Rightarrow$

$d_{ij}=0$ ($i=1,\ldots,n; j=1,\ldots,n; j\neq i$), while $r=0.5$;

Because $d_{i0}=0$ and $d_{ij}=0$ ($i=1,\ldots,n; j=1,\ldots,n; j\neq i$) while $r=0.5$, $U(X,0.5,r0)=0$.

Thus, a fake sphere configuration $X$ whose $U(X,0.5+10^{-8},r0)<10^{-16}$ is an exact packing $X$ of standard spheres whose $U(X,0.5,r0)=0$.

The above theorem can be illustrated in Fig.1.

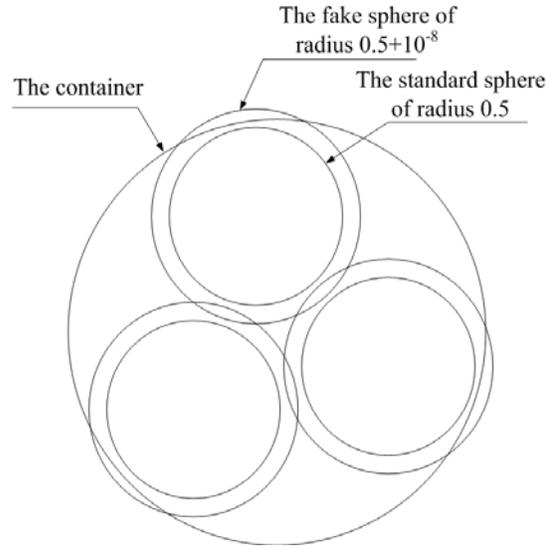

Figure 1: The fake sphere trick.

Henceforth, given a configuration $X$ of $n$ equal spheres with radius $r$, we denote the process that A0 develops $X$ into $X_{halt}$ inside a container of radius $r0$, by $X_{halt} \leftarrow A0(r,r0,X)$.

**The serial symmetrical relocation strategy**

Given a local optimal configuration $X_{local}$ of $n$ equal spheres, any slight disturbance in it can not change its spatial structure and is unlikely to improve it. But if we move the $i^{th}$ sphere in this configuration from its current location $(x_i,y_i,z_i)$ to $(-x_i,-y_i,-z_i)$ and keep other spheres unmoved, the spatial structure of $X_{local}$ must be greatly changed.

**Definition (symmetrical relocation)** *Given a configuration X, the symmetrical relocation of the $i^{th}$ sphere of X is the action of changing the $i^{th}$ sphere center from $x_i,y_i,z_i$ to $-x_i,-y_i,-z_i$.*

The following denotations are introduced to describe the serial symmetrical relocation strategy.

$X_{local}$ stands for a local optimal configuration which is viewed as a set of the locations of $n$ equal spheres centers.

$minU(i,L)$ stands for the $i$ spheres with the smallest potential energies among all spheres in set $L$. $L$ is a subset of $X_{local}$.

$maxU(i,L)$ stands for the $i$ spheres with the largest potential energies among all spheres in set $L$. $L$ is a subset of $X_{local}$.

$invert(L,X_{local})$ stands for the new configuration obtained by symmetrically relocating all spheres in the subset $L$ of $X_{local}$ and keeping all other spheres in $X_{local}$ unmoved.

The serial symmetrical relocation strategy is to generate $n(n-1)/2$ configurations from a local optimal configuration $X_{local}$ of $n$ spheres by

$$invert(maxU(j,minU(i,X_{local})),X_{local}), 1 \leq i \leq n,\ 1 \leq j < i \qquad (6)$$

Since the configuration $invert(maxU(n-i,minU(n,X_{local})),X_{local})$ is identical to the configuration $invert(maxU(i,minU(i,X_{local})),X_{local})$ for symmetry, the variable $j$ ranges from 1 to $i$-1. The number of the new configurations obtained by (6) is $n(n-1)/2$.

**The serial symmetrical relocation algorithm**

Here, we introduce the serial symmetrical relocation algorithm which is denoted by A1. The inputs of A1 are a roughly estimated value of container radius $r0$ and a randomly generated configuration. The outputs are a dense packing and its exact container radius.

The core of A1 is the heuristic strategy: serial symmetrical relocation strategy. In A1, the serial symmetrical relocation strategy each time generates $n(n-1)/2$ new configurations from a local optimal configuration $X_{local}$ and calls A0 to examine them one by one. The process of generating $n(n-1)/2$ new configurations from one $X_{local}$ and examining them by A0 is called one scan of A1. The first $X_{local}$ which starts the first scan of A1 is obtained from a randomly generated initial configuration. Each scan of A1, except the first scan, takes the configuration with the smallest potential energy found in its previous scan as its $X_{local}$.

The pseudo code of A1 is given as follows:
/* The beginning of pseudo code of A1*/
$n \leftarrow$ the number of equal spheres;

$r0 \leftarrow$ an empirically estimated value of container radius for *n* sphere;

$r \leftarrow 0.5+10^{-8}$; /* The fake sphere trick. */

Generates one random initial configuration $X_{random}$ of *n* spheres;

/* From a randomly generated configuration $X_{random}$, we get the first $X_{local}$. */

$X_{local} \leftarrow A0(r,r0,X_{random})$;

if ($U(X_{local},r,r0)<10^{-16}$) {

   $X_{found} \leftarrow X_{local}$;

   Goto Binary_Search; } /* A1 finds a packing.*/

$l \leftarrow 0$; /* Scan counter. */

while ($l< 6$) /* 6 is a manually predetermined upper limit of scan number. */

{

   for *i*=1 to *n*

   {

      for *j*=1 to *i*-1

      {

         $X_{newlocal}^{(ij)} \leftarrow A0(r,r0, invert(maxU(j,minU(i,X_{local})),X_{local}))$ ;

         if ($U(X_{newlocal}^{(ij)},r,r0)<10^{-16}$) {

            $X_{found} \leftarrow X_{newlocal}^{(ij)}$;

            Goto Binary_Search; } /* A1 finds a packing. */

      }

   }

   $X_{local} \leftarrow$ the one with the lowest potential energy among all $X_{newlocal}^{(ij)}$;

   $l \leftarrow l+1$;

} /* End of while*/

/* If A1 doesn't directly find any packing, pick the best configuration ever found.*/

$X_{found} \leftarrow$ the local optimal configuration with the lowest potential energy ever found during all scans;

Binary_Search:

/* Use a binary search to find the smallest container in which $X_{found}$ can just be developed by A0 into a packing. */

$r0_{low} \leftarrow 1/2 \times r0$;

$r0_{up} \leftarrow 2 \times r0$;

$\varepsilon \leftarrow 10^{-12}$;  /* $\varepsilon$ is a precision. */

$X \leftarrow X_{found}$;

while ($r0_{up} - r0_{low} > \varepsilon$)

{

    $r0 \leftarrow (r0_{up} + r0_{low})/2$;

    $X \leftarrow A0(r, r0, X)$;

    if ($U(X, r, r0) < 10^{-16}$)

        $r0_{up} \leftarrow r0$;

    if ($U(X, r, r0) \geq 10^{-16}$)

        $r0_{low} \leftarrow r0$;

}

$r0_{min} \leftarrow r0_{up}$;

/* $r0_{min}$ is the smallest container radius for $X_{found}$ to become a packing through A0.*/

$X_{dense} \leftarrow A0(r, r0_{min}, X)$;

/* $X_{dense}$ is the corresponding dense packing of $X_{found}$ in the container of $r0_{min}$.*/

/* The end of pseudo code of A1*/

A1 examines only $O(n^2)$ configurations, thus it can evaluate the instances of up to 200 spheres within feasible runtime.

For each instance, we performed A1 five times and got five dense packings. We chose the best one of them as the result of the instance.

**Numerical results**

We performed A1 on a personal computer with Pentium E6500 2.93GHz and 2GB DDR2 800MHZ RAM and obtained dense packings of equal spheres in spherical and cube containers. All our packings are rigorous because of the fake sphere trick.

Because of the basic nature of the equal sphere packing problem, seemingly "insignificant" improvements on the former records may imply new packings which are unknown before. The nature of the equal sphere packing problem is analogous to that of the equal circle packing problem. (Specht, 2012) collects the best known records of the equal circle packing problem from different researchers. Among these best known records, (Buddenhagen, 2010) reduced the circumcircle radius of 64 unit circles from the former record 8.96197110850392353216121 to 8.96197110848573830216130. This is a real improvement. Thus, according to the quality of the results found by it, we assume the serial symmetrical relocation algorithm (denoted by A1) lead to substantial improvements.

Especially, the serial symmetrical relocation algorithm packed 68 equal spheres of radius 1 into a large sphere of radius 1/0.20000222, and proved wrong a conjecture (Pfoertner, 2011) which alleges a large sphere of radius 5 can contain at most 67 equal spheres of radius 1.

**The results for packing equal spheres inside spherical containers**

We list in Table 1 the containers sizes for our best packings in spherical containers, along with the best known records to our knowledge. The quality of spherical container is represented by the ratio $r/r0$, where $r$ is 0.5.

Table 1. Records for packing equal spheres in spherical containers.

| $n$ | A1 $r/r0$ | Average runtime for A1 $r/r0$ (second) | Hugo(a) $r/r0$ | Hugo(b) $r/r0$ | Dave $r/r0$ | A1 $r/r0$ minus Hugo(b) $r/r0$ |
|---|---|---|---|---|---|---|
| 1 | 1.00000000 | 0.001 | 1.00000000 | 1.00000000 | - | 0.00000000 |
| 2 | 0.50000000 | 0.001 | 0.50000000 | 0.50000000 | - | 0.00000000 |
| 3 | 0.46410160 | 0.006 | 0.46410160 | 0.46410160 | - | 0.00000000 |
| 4 | 0.44948974 | 0.007 | 0.44948970$^{(+)}$ | 0.44948970$^{(+)}$ | - | 0.00000004 |
| 5 | 0.41421350 | 0.025 | 0.41421350 | 0.41421350 | - | 0.00000000 |
| 6 | 0.41421350 | 0.043 | 0.41421350 | 0.41421350 | - | 0.00000000 |
| 7 | 0.38591355 | 0.079 | 0.38591350$^{(+)}$ | 0.38591360$^{(-)}$ | - | -0.00000005 |
| 8 | 0.37802480 | 0.134 | 0.37802480 | 0.37802480 | - | 0.00000000 |
| 9 | 0.36602539 | 1.216 | 0.36602530$^{(+)}$ | 0.36602540$^{(-)}$ | - | -0.00000001 |
| 10 | 0.35304942 | 0.926 | 0.35304940$^{(+)}$ | 0.35304940$^{(+)}$ | - | 0.00000002 |
| 11 | 0.34457650 | 0.776 | 0.34457640$^{(+)}$ | 0.34457650 | - | 0.00000000 |
| 12 | 0.34457650 | 0.675 | 0.34457640$^{(+)}$ | 0.34457650 | - | 0.00000000 |
| 13 | 0.33333332 | 1.028 | 0.33333330 | 0.33333330 | - | 0.00000002 |
| 14 | 0.32350466 | 1.148 | 0.32350460$^{(+)}$ | 0.32350460$^{(+)}$ | 0.32331300$^{(+)}$ | 0.00000006 |
| 15 | 0.31830481 | 1.345 | 0.31830470$^{(+)}$ | 0.31830480$^{(+)}$ | 0.31830500$^{(-)}$ | 0.00000001 |
| 16 | 0.31097591 | 2.330 | 0.31097580$^{(+)}$ | 0.31097590$^{(+)}$ | 0.31097600$^{(-)}$ | 0.00000001 |
| 17 | 0.30569395 | 2.914 | 0.30569390$^{(+)}$ | 0.30569400$^{(-)}$ | 0.30569400$^{(-)}$ | -0.00000005 |
| 18 | 0.30129658 | 3.016 | 0.30129650$^{(+)}$ | 0.30129650$^{(+)}$ | 0.30129600$^{(-)}$ | 0.00000008 |
| 19 | 0.29533232 | 4.635 | 0.29533230$^{(+)}$ | 0.29533230$^{(+)}$ | 0.29533200$^{(+)}$ | 0.00000002 |
| 20 | 0.28789082 | 5.678 | 0.28789070$^{(+)}$ | 0.28789070$^{(+)}$ | 0.28785100$^{(+)}$ | 0.00000012 |
| 21 | 0.28683281 | 8.321 | 0.28683270$^{(+)}$ | 0.28683270$^{(+)}$ | 0.28683300$^{(-)}$ | 0.00000011 |
| 22 | 0.27934262 | 6.916 | 0.27934250$^{(+)}$ | 0.27934250$^{(+)}$ | 0.27933400$^{(+)}$ | 0.00000012 |
| 23 | 0.27567069 | 9.199 | 0.27567070$^{(-)}$ | 0.27567070$^{(-)}$ | 0.27508100$^{(+)}$ | -0.00000001 |
| 24 | 0.27134130 | 10.783 | 0.27134130 | 0.27134130 | 0.27133600$^{(+)}$ | 0.00000000 |
| 25 | 0.27119182 | 13.595 | 0.27119170$^{(+)}$ | 0.27119180$^{(+)}$ | 0.27112000$^{(+)}$ | 0.00000002 |
| 26 | 0.26685126 | 14.057 | 0.26684960$^{(+)}$ | 0.26685130$^{(-)}$ | 0.26667920$^{(+)}$ | -0.00000004 |
| 27 | 0.26223207 | 17.722 | 0.26223200$^{(+)}$ | 0.26223210$^{(-)}$ | 0.26212000$^{(+)}$ | -0.00000003 |
| 28 | 0.26030547 | 21.977 | 0.26030080$^{(+)}$ | 0.26030550$^{(-)}$ | 0.26009600$^{(+)}$ | -0.00000003 |
| 29 | 0.25792545 | 35.987 | 0.25792510$^{(+)}$ | 0.25792550$^{(-)}$ | 0.25781900$^{(+)}$ | -0.00000005 |
| 30 | 0.25533055 | 26.826 | 0.25533000$^{(+)}$ | 0.25533060$^{(-)}$ | 0.25478000$^{(+)}$ | -0.00000005 |
| 31 | 0.25311620 | 79.014 | 0.25311410$^{(+)}$ | 0.25311620 | 0.25311500$^{(+)}$ | 0.00000000 |
| 32 | 0.25078744 | 38.078 | 0.25078740$^{(+)}$ | 0.25078740$^{(+)}$ | 0.25071200$^{(+)}$ | 0.00000004 |
| 33 | 0.24876230 | 39.542 | 0.24871040$^{(+)}$ | 0.24876240$^{(+)}$ | 0.24870300$^{(+)}$ | -0.00000010 |
| 34 | 0.24705265 | 43.430 | 0.24704220$^{(+)}$ | 0.24705270$^{(-)}$ | 0.24700600$^{(+)}$ | -0.00000005 |
| 35 | 0.24483365 | 47.870 | 0.24482840$^{(+)}$ | 0.24483370$^{(-)}$ | 0.24477300$^{(+)}$ | -0.00000005 |
| 36 | 0.24313216 | 56.587 | 0.24313210$^{(+)}$ | 0.24313220$^{(-)}$ | 0.24178700$^{(+)}$ | -0.00000004 |
| 37 | 0.24068655 | 61.810 | 0.24062170$^{(+)}$ | 0.24068660$^{(-)}$ | 0.24044100$^{(+)}$ | -0.00000005 |
| 38 | 0.24051936 | 98.767 | 0.24051470$^{(+)}$ | 0.24051560$^{(+)}$ | 0.24036000$^{(+)}$ | 0.00000376 |
| 39 | 0.23674523 | 77.186 | 0.23655210$^{(+)}$ | 0.23674520$^{(+)}$ | 0.23670300$^{(+)}$ | 0.00000003 |
| 40 | 0.23499923 | 85.288 | 0.23497560$^{(+)}$ | 0.23499920$^{(+)}$ | 0.23487100$^{(+)}$ | 0.00000003 |
| 41 | 0.23275597 | 92.837 | 0.23266430$^{(+)}$ | 0.23275590$^{(+)}$ | 0.23270200$^{(+)}$ | 0.00000007 |
| 42 | 0.23211860 | 109.130 | 0.23184150$^{(+)}$ | 0.23211900$^{(-)}$ | 0.23211900$^{(-)}$ | -0.00000040 |
| 43 | 0.22973295 | 113.118 | 0.22934050$^{(+)}$ | 0.22973300$^{(-)}$ | 0.22966700$^{(+)}$ | -0.00000005 |
| 44 | 0.22816302 | 128.813 | 0.22815860$^{(+)}$ | 0.22816290$^{(+)}$ | 0.22808300$^{(+)}$ | 0.00000012 |
| 45 | 0.22691155 | 233.270 | 0.22674120$^{(+)}$ | 0.22691160$^{(-)}$ | 0.22659400$^{(+)}$ | -0.00000005 |
| 46 | 0.22516818 | 245.517 | 0.22511610$^{(+)}$ | 0.22516810$^{(+)}$ | 0.22516400$^{(+)}$ | 0.00000008 |
| 47 | 0.22350704 | 148.588 | 0.22348200$^{(+)}$ | 0.22350700$^{(+)}$ | 0.22334200$^{(+)}$ | 0.00000004 |
| 48 | 0.22240593 | 178.593 | 0.22235400$^{(+)}$ | 0.22240590$^{(+)}$ | 0.22236300$^{(+)}$ | 0.00000003 |
| 49 | 0.22127817 | 188.396 | 0.22091760$^{(+)}$ | 0.22127510$^{(+)}$ | 0.22098000$^{(+)}$ | 0.00000307 |
| 50 | 0.21975827 | 233.125 | 0.21934840$^{(+)}$ | 0.21975290$^{(+)}$ | - | 0.00000537 |
| 51 | 0.21855027 | 229.869 | 0.21788800$^{(+)}$ | 0.21788880$^{(+)}$ | - | 0.00066147 |
| 52 | 0.21693040 | 287.637 | - | 0.21676650$^{(+)}$ | - | 0.00016390 |
| 53 | 0.21628805 | 259.448 | - | 0.21577720$^{(+)}$ | - | 0.00051085 |
| 54 | 0.21492066 | 296.349 | - | 0.21468460$^{(+)}$ | - | 0.00023606 |
| 55 | 0.21344168 | 297.398 | - | 0.21326800$^{(+)}$ | - | 0.00017368 |
| 56 | 0.21306786 | 379.628 | - | 0.21261360$^{(+)}$ | - | 0.00045426 |
| 57 | 0.21130935 | 340.078 | - | 0.21086610[(+)] | - | 0.00044325 |
| 58 | 0.21048009 | 397.783 | - | 0.21030220[(+)] | - | 0.00017789 |
| 59 | 0.20975944 | 391.824 | - | 0.20962520[(+)] | - | 0.00013424 |
| 60 | 0.20942699 | 483.227 | - | 0.20942690[(+)] | - | 0.00000009 |
| 61 | 0.20910865 | 412.989 | - | 0.20891230[(+)] | - | 0.00019635 |
| 62 | 0.20673915 | 452.234 | - | 0.20581700[(+)] | - | 0.00092215 |
| 63 | 0.20595491 | 519.972 | - | 0.20471400[(+)] | - | 0.00124091 |
| 64 | 0.20408715 | 685.259 | - | 0.20366460[(+)] | - | 0.00042255 |
| 65 | 0.20307603 | 589.248 | - | 0.20274750[(+)] | - | 0.00032853 |
| 66 | 0.20210875 | 636.278 | - | 0.20191580[(+)] | - | 0.00019295 |
| 67 | 0.20121662 | 854.971 | - | 0.20044290[(+)] | - | 0.00077372 |
| 68 | 0.20000222 | 694.894 | - | 0.19972200[(+)] | - | 0.00028022 |
| 69 | 0.19928152 | 706.689 | - | 0.19879220[(+)] | - | 0.00048932 |
| 70 | 0.19868872 | 786.326 | - | 0.19833100[(+)] | - | 0.00035772 |
| 71 | 0.19722212 | 805.879 | - | 0.19697870[(+)] | - | 0.00024342 |
| 72 | 0.19628591 | 795.408 | - | 0.19580010[(+)] | - | 0.00048581 |
| 73 | 0.19562734 | 992.946 | - | - | - | - |
| 74 | 0.19514678 | 1074.563 | - | - | - | - |
| 75 | 0.19396442 | 1068.320 | - | - | - | - |
| 76 | 0.19294742 | 1183.277 | - | - | - | - |
| 77 | 0.19225411 | 1243.473 | - | - | - | - |
| 78 | 0.19145892 | 1208.989 | - | - | - | - |
| 79 | 0.19065973 | 1229.879 | - | - | - | - |
| 80 | 0.18972728 | 1336.218 | - | - | - | - |
| 81 | 0.18896712 | 1493.084 | - | - | - | - |
| 82 | 0.18829304 | 1452.398 | - | - | - | - |
| 83 | 0.18764017 | 1592.395 | - | - | - | - |
| 84 | 0.18695602 | 1608.082 | - | - | - | - |
| 85 | 0.18641122 | 1791.520 | - | - | - | - |
| 86 | 0.18561006 | 1776.278 | - | - | - | - |
| 87 | 0.18495292 | 1809.917 | - | - | - | - |
| 88 | 0.18427588 | 1849.809 | - | - | - | - |
| 89 | 0.18348126 | 2239.128 | - | - | - | - |
| 90 | 0.18293732 | 2102.385 | - | - | - | - |
| 91 | 0.18235327 | 2848.673 | - | - | - | - |
| 92 | 0.18177412 | 2307.977 | - | - | - | - |
| 93 | 0.18108200 | 2331.690 | - | - | - | - |
| 94 | 0.18047196 | 2537.434 | - | - | - | - |
| 95 | 0.18011439 | 2847.453 | - | - | - | - |
| 96 | 0.17956685 | 2660.676 | - | - | - | - |
| 97 | 0.17893218 | 2807.216 | - | - | - | - |
| 98 | 0.17830007 | 2999.697 | - | - | - | - |
| 99 | 0.17787965 | 3021.932 | - | - | - | - |
| 100 | 0.17743921 | 3378.980 | - | - | - | - |
| 105 | 0.17459442 | 3902.393 | - | - | - | - |
| 110 | 0.17212934 | 4958.305 | - | - | - | - |
| 115 | 0.16970936 | 5782.270 | - | - | - | - |
| 120 | 0.16749907 | 6789.297 | - | - | - | - |
| 125 | 0.16543987 | 5985.572 | - | - | - | - |
| 130 | 0.16373513 | 9786.393 | - | - | - | - |
| 135 | 0.16210499 | 12794.897 | - | - | - | - |
| 140 | 0.16029921 | 15986.278 | - | - | - | - |
| 145 | 0.15859944 | 12363.627 | - | - | - | - |
| 150 | 0.15665832 | 16893.134 | - | - | - | - |
| 155 | 0.15470631 | 19769.023 | - | - | - | - |
| 160 | 0.15335971 | 20689.204 | - | - | - | - |
| 165 | 0.15178309 | 25019.026 | - | - | - | - |
| 170 | 0.15034579 | 22134.377 | - | - | - | - |
| 175 | 0.14889938 | 31281.456 | - | - | - | - |
| 180 | 0.14741802 | 34897.343 | - | - | - | - |
| 185 | 0.14604949 | 38968.805 | - | - | - | - |
| 190 | 0.14445529 | 40354.659 | - | - | - | - |
| 195 | 0.14337725 | 39691.728 | - | - | - | - |
| 200 | 0.14224761 | 50982.358 | - | - | - | - |

In Table 1, a superscript (+) of one record indicates that the corresponding record of A1 $r/r0$ is larger (better) than this record. For example, the record for 72 equal spheres in Hugo(b) $r/r0$ has a superscript (+), because the corresponding record of A1 $r/r0$ is larger than it by $4.8581 \times 10^{-4}$. The superscript (-) has the opposite meaning.

In Table 1, A1 $r/r0$ is obtained by the serial symmetrical relocation algorithm. Hugo(a) $r/r0$ is a set of verifiable records provided by Hugo Pfoertner (Pfoertner, 2008a), which are supported by their coordinates of spheres centers. Hugo(b) $r/r0$ is a set of best known records collected by Hugo Pfoertner (Pfoertner, 2008b), which is produced by Hugo Pfoertner, Thierry Gensane and Dave Boll. Each record of this collection is the best one of their respective records. Dave $r/r0$ is provided by Dave Boll (Boll, 2005).

Some packings found by the serial symmetrical relocation algorithm inside spherical containers are illustrated in Fig.2.

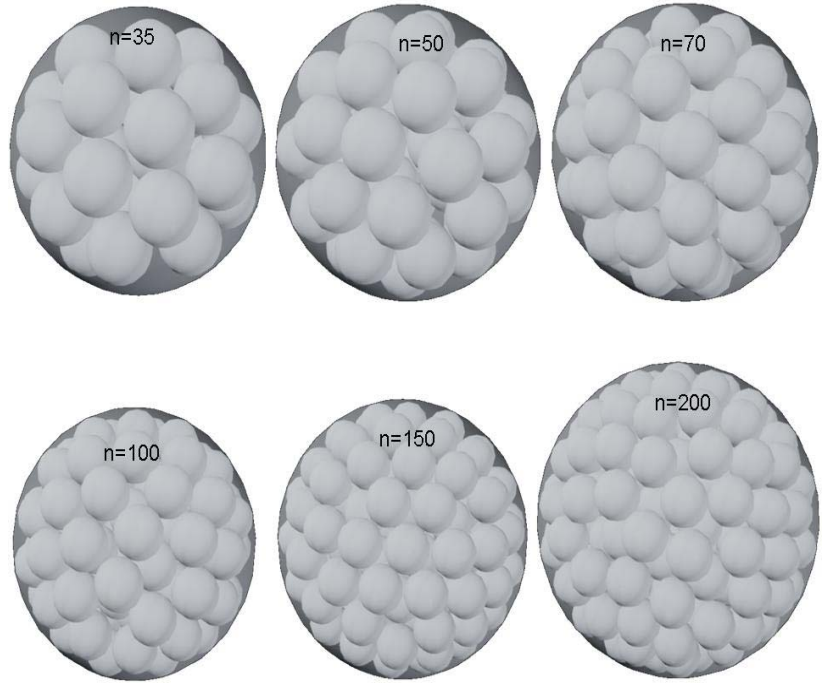

Figure 2: Six packings of equal spheres inside sphere

**The results for packing equal spheres inside cubes**

To test the generality of A1, we have also packed equal spheres inside cubes. The cubes sizes of our best packings are listed in table 2, along with the best known records to our knowledge. Here, we take one half of cube edge as cube radius $r_0$. The quality of cube is represented by the ratio $r/r_0$, where $r$ is 0.5.

Table 2. Records for packing equal spheres in cubes.

| $n$ | A1 $r/r_0$ | Average runtime for A1 $r/r_0$ (second) | Hugo $r/r_0$ | A1 $r/r_0$ minus Hugo $r/r_0$ |
|---|---|---|---|---|
| 1 | 1.00000000 | 0.001 | 1.00000000 | 0.00000000 |
| 2 | 0.63397458 | 0.001 | 0.63397460$^{(-)}$ | -0.00000002 |
| 3 | 0.58578640 | 0.008 | 0.58578640 | 0.00000000 |
| 4 | 0.58578640 | 0.006 | 0.58578640 | 0.00000000 |
| 5 | 0.52786400 | 0.032 | 0.52786400 | 0.00000000 |
| 6 | 0.51471861 | 0.033 | 0.51471860$^{(+)}$ | 0.00000001 |
| 7 | 0.50027231 | 0.058 | 0.50027230$^{(+)}$ | 0.00000001 |
| 8 | 0.50000000 | 0.182 | 0.50000000 | 0.00000000 |
| 9 | 0.46410161 | 1.016 | 0.46410160$^{(+)}$ | 0.00000001 |
| 10 | 0.42857142 | 1.126 | 0.42857140$^{(+)}$ | 0.00000002 |
| 11 | 0.41524447 | 0.976 | 0.41524450$^{(-)}$ | -0.00000003 |
| 12 | 0.41421355 | 0.682 | 0.41421350$^{(+)}$ | 0.00000005 |
| 13 | 0.41421355 | 1.223 | 0.41421350$^{(+)}$ | 0.00000005 |
| 14 | 0.41421355 | 1.198 | 0.41421350$^{(+)}$ | 0.00000005 |
| 15 | 0.38461537 | 1.648 | 0.38461540$^{(-)}$ | -0.00000003 |
| 16 | 0.37759353 | 2.832 | 0.37759350$^{(+)}$ | 0.00000003 |
| 17 | 0.37737047 | 2.936 | 0.37737050$^{(-)}$ | -0.00000003 |
| 18 | 0.37536119 | 3.523 | 0.37536110$^{(+)}$ | 0.00000009 |
| 19 | 0.36637060 | 4.358 | 0.36637040$^{(+)}$ | 0.00000020 |
| 20 | 0.35681439 | 5.925 | 0.35681440$^{(-)}$ | -0.00000001 |
| 21 | 0.35443808 | 7.368 | 0.35443810$^{(-)}$ | -0.00000002 |
| 22 | 0.34654620 | 9.932 | 0.34654620 | 0.00000000 |
| 23 | 0.34363356 | 8.996 | 0.34363330$^{(+)}$ | 0.00000026 |
| 24 | 0.34108137 | 11.653 | 0.34108110$^{(+)}$ | 0.00000027 |
| 25 | 0.33560809 | 12.893 | 0.33560750$^{(+)}$ | 0.00000059 |
| 26 | 0.33381032 | 15.959 | 0.33381020$^{(+)}$ | 0.00000012 |
| 27 | 0.33333332 | 20.732 | 0.33333330$^{(+)}$ | 0.00000002 |
| 28 | 0.32038200 | 22.803 | 0.32038200 | 0.00000000 |
| 29 | 0.32037723 | 23.682 | 0.32037720$^{(+)}$ | 0.00000003 |
| 30 | 0.32037723 | 27.376 | 0.32037720$^{(+)}$ | 0.00000003 |
| 31 | 0.32037723 | 33.518 | 0.32037720$^{(+)}$ | 0.00000003 |
| 32 | 0.32037723 | 35.142 | 0.32037720$^{(+)}$ | 0.00000003 |
| 33 | 0.30921071 | 38.252 | 0.30921070$^{(+)}$ | 0.00000001 |
| 34 | 0.30423418 | 33.426 | 0.30423380$^{(+)}$ | 0.00000038 |
| 35 | 0.30333704 | 46.853 | 0.30333690$^{(+)}$ | 0.00000014 |
| 36 | 0.29861826 | 57.087 | 0.29861740$^{(+)}$ | 0.00000086 |
| 37 | 0.29812325 | 66.823 | 0.29812320$^{(+)}$ | 0.00000005 |
| 38 | 0.29807753 | 68.367 | 0.29807750$^{(+)}$ | 0.00000003 |
| 39 | 0.29523475 | 78.236 | 0.29523470$^{(+)}$ | 0.00000005 |
| 40 | 0.29411764 | 95.218 | 0.29411760$^{(+)}$ | 0.00000004 |
| 41 | 0.28967575 | 91.825 | 0.28967510$^{(+)}$ | 0.00000065 |
| 42 | 0.28608925 | 107.653 | 0.28608530$^{(+)}$ | 0.00000395 |
| 43 | 0.28330422 | 115.127 | 0.28330320$^{(+)}$ | 0.00000102 |
| 44 | 0.28172121 | 118.423 | 0.28172060$^{(+)}$ | 0.00000061 |
| 45 | 0.28126386 | 133.370 | 0.28126390$^{(-)}$ | -0.00000004 |
| 46 | 0.28049373 | 143.527 | 0.28049340$^{(+)}$ | 0.00000033 |
| 47 | 0.27991768 | 146.982 | 0.27991770$^{(-)}$ | -0.00000002 |
| 48 | 0.27991768 | 288.125 | 0.27991770$^{(-)}$ | -0.00000002 |
| 49 | 0.27285349 | 198.592 | 0.27285340$^{(+)}$ | 0.00000009 |
| 50 | 0.27190872 | 225.216 | 0.27190740$^{(+)}$ | 0.00000132 |
| 51 | 0.27004592 | 236.169 | 0.27001290$^{(+)}$ | 0.00003302 |
| 52 | 0.26764001 | 272.237 | 0.26763980$^{(+)}$ | 0.00000021 |
| 53 | 0.26646158 | 236.426 | 0.26646040$^{(+)}$ | 0.00000118 |
| 54 | 0.26330123 | 299.028 | 0.26326990$^{(+)}$ | 0.00003133 |
| 55 | 0.26214603 | 293.256 | 0.26176990$^{(+)}$ | 0.00037613 |
| 56 | 0.26130505 | 279.338 | 0.26121580$^{(+)}$ | 0.00008925 |
| 57 | 0.26120393 | 306.120 | 0.26120380$^{(+)}$ | 0.00000013 |
| 58 | 0.26120387 | 396.085 | 0.26120380$^{(+)}$ | 0.00000007 |
| 59 | 0.26120387 | 386.516 | 0.26120380$^{(+)}$ | 0.00000007 |
| 60 | 0.26120387 | 459.205 | 0.26120380$^{(+)}$ | 0.00000007 |
| 61 | 0.26120387 | 412.989 | 0.26120390$^{(-)}$ | -0.00000003 |
| 62 | 0.26120387 | 483.122 | 0.26120380$^{(+)}$ | 0.00000007 |
| 63 | 0.26120387 | 536.268 | 0.26120380$^{(+)}$ | 0.00000007 |
| 64 | 0.25348654 | 583.427 | 0.25348040$^{(+)}$ | 0.00000614 |
| 65 | 0.25236517 | 589.248 | 0.25235460$^{(+)}$ | 0.00001057 |
| 66 | 0.25147929 | 616.163 | 0.25146920$^{(+)}$ | 0.00001009 |
| 67 | 0.24896216 | 653.073 | 0.24896000$^{(+)}$ | 0.00000216 |
| 68 | 0.24686760 | 687.253 | 0.24675700$^{(+)}$ | 0.00011060 |
| 69 | 0.24627801 | 702.585 | 0.24593500$^{(+)}$ | 0.00034301 |
| 70 | 0.24607500 | 883.166 | 0.24588280$^{(+)}$ | 0.00019220 |
| 71 | 0.24605515 | 805.879 | 0.24569490$^{(+)}$ | 0.00036025 |
| 72 | 0.24605515 | 795.408 | 0.24566920$^{(+)}$ | 0.00038595 |
| 73 | 0.24357499 | 998.258 | - | - |
| 74 | 0.24267720 | 1182.591 | - | - |
| 75 | 0.24264288 | 1008.150 | - | - |
| 76 | 0.24036710 | 1283.368 | - | - |
| 77 | 0.23810069 | 1353.386 | - | - |
| 78 | 0.23709060 | 1198.385 | - | - |
| 79 | 0.23602798 | 1329.125 | - | - |
| 80 | 0.23483879 | 1306.856 | - | - |
| 81 | 0.23414746 | 1503.165 | - | - |
| 82 | 0.23366493 | 1497.893 | - | - |
| 83 | 0.23256887 | 1539.263 | - | - |
| 84 | 0.23256724 | 1678.925 | - | - |
| 85 | 0.23256724 | 2098.535 | - | - |
| 86 | 0.23256724 | 1769.892 | - | - |
| 87 | 0.23256724 | 1822.735 | - | - |
| 88 | 0.23104286 | 1856.203 | - | - |
| 89 | 0.22753889 | 2369.213 | - | - |
| 90 | 0.22651347 | 2086.563 | - | - |
| 91 | 0.22558935 | 3003.356 | - | - |
| 92 | 0.22500279 | 2687.076 | - | - |
| 93 | 0.22453635 | 2030.548 | - | - |
| 94 | 0.22398206 | 2612.231 | - | - |

| 95 | 0.22347071 | 2896.546 | - | - |
| 96 | 0.22329903 | 2698.323 | - | - |
| 97 | 0.22309749 | 2936.285 | - | - |
| 98 | 0.22303798 | 2597.805 | - | - |
| 99 | 0.22277482 | 3136.025 | - | - |
| 100 | 0.22276469 | 3295.895 | - | - |
| 105 | 0.22048120 | 3979.295 | - | - |
| 110 | 0.21470034 | 5085.573 | - | - |
| 115 | 0.21027862 | 5968.735 | - | - |
| 120 | 0.20959841 | 7937.739 | - | - |
| 125 | 0.20710709 | 8283.382 | - | - |
| 130 | 0.20297413 | 10063.240 | - | - |
| 135 | 0.20005504 | 11373.453 | - | - |
| 140 | 0.19972284 | 16901.682 | - | - |
| 145 | 0.19642072 | 18313.383 | - | - |
| 150 | 0.19339963 | 15293.625 | - | - |

In Table 2, the superscripts (+) and (-) have the same meanings as before. A1 *r/r0* is obtained by the serial symmetrical relocation algorithm. Hugo *r/r0* is a collection of records gathered by Hugo Pfoertner (Pfoertner, 2005), which is produced by J. Schaer, M. Goldberg, Hugo Pfoertner, Thierry Gensane and Dave Boll.

Some packings found by the serial symmetrical relocation algorithm inside cubes are illustrated in Fig.3.

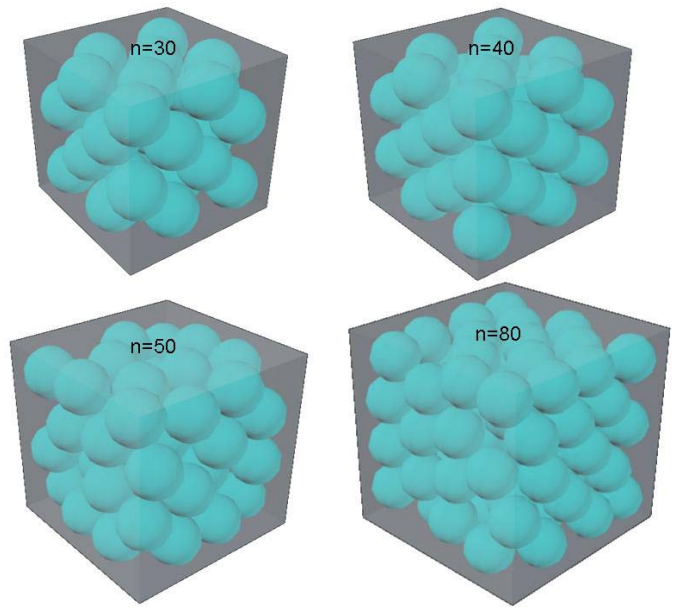

Figure 3: Four packings of equal spheres inside cube

## Conclusions

The serial symmetrical relocation strategy as the global search strategy is the core of our work. Experiments show that it led to some improvements over the current best known records about the equal sphere packing problem.

The fake sphere trick guarantees that all results found by the serial symmetrical relocation algorithm are rigorous. We stress the exactness of results because any approximate result including overlaps may have the exaggerated quality and, thus may cover any possible subtle improvements.

By intuition, the serial symmetrical relocation algorithm could be applied for packing circles or spheres in any centrosymmetric container. We will further develop the ideas that presented in this paper and try to find out a generic and highly efficient method for packing circles or spheres in arbitrary containers.

## Acknowledgements

The authors thank Hugo et al for their records. The authors also thank anonymous referees for their helpful comments and suggestions.